\documentclass[12pt]{article}
\usepackage{a4,epsf}

\newcommand \ket[1]{\left\vert\, {#1} \, \right>}

\newcommand{\bea}{\begin{eqnarray}}
\newcommand{\eea}{\end{eqnarray}}
\newcommand{\simgt}{\hbox{ \raise3pt\hbox to 0pt{$>$}\raise-3pt\hbox{$\sim$} }}
\newcommand{\simlt}{\hbox{ \raise3pt\hbox to 0pt{$<$}\raise-3pt\hbox{$\sim$} }}

\begin{document}
\begin{titlepage}
\title{
\vspace{2cm}
Neutrino mixing and see-saw mechanism}
\author{M.~Je\.zabek$^{a,b)}$
and
Y.~Sumino$^{c)}$\thanks{On 
leave of absence from Department of Physics, Tohoku University,
Sendai 980-77, Japan.}
\\ \\ \\ \small
$a)$ Institute of Nuclear Physics,
Kawiory 26a, PL-30055 Cracow, Poland
\\   \small
  $b)$ Department of Field Theory and Particle Physics, University of 
      Silesia, \\[-5pt]  \small
     Uniwersytecka 4, PL-40007 Katowice, Poland\\ 
\small
$c)$ Institut f\"ur Theoretische Teilchenphysik,
Universit\"at Karlsruhe,\\  [-5pt]
\small
D-76128 Karlsruhe, Germany
}
\date{}
\maketitle
\thispagestyle{empty}
\vspace{-4.5truein}
\begin{flushright}
{\bf TP-USl/98/6}\\
{\bf TTP 98-26}\\
{\bf hep-ph/9807310}\\
{\bf July 1998}
\end{flushright}
\vspace{4.0truein}
\begin{abstract}
Models of neutrino masses are discussed capable of explaining
in a natural way the maximal mixing between $\nu_{\mu}$ and
$\nu_{\tau}$ observed by the Super-Kamiokande collaboration. 
For three generations of leptons two
classes of such models are found implying: \\
a) 
$\Delta {m_{23}}^2 \ll \Delta {m_{12}}^2 \approx \Delta {m_{13}}^2 $
and a small mixing between $\nu_{e}$ and the other two neutrinos,\\
b) 
$\Delta {m_{12}}^2 \ll \Delta {m_{13}}^2 \approx \Delta {m_{23}}^2$
and a nearly maximal mixing for solar neutrino oscillations in vacuum.
\end{abstract}
\end{titlepage}

The recent results from Super-Kamiokande \cite{SKam}  indicate a large
mixing of $\nu_\mu$ and $\nu_\tau$. In fact the observed mixing angle is 
nearly maximal
\begin{equation}
\sin^2\, 2\theta_{\mu\tau} \approx 1 .
\label{mutaumix}
\end{equation}
In this note we study a problem if this remarkably large mixing can be
explained in a natural way. We assume that the masses of neutrinos are
much smaller than the masses of quarks and charged leptons due to the
see-saw mechanism \cite{YGMRS}, see also an excellent review \cite{FY}.  
For three generations of neutrinos the mass matrix is of the form
\begin{equation}
{\bf M_{6}} \, = \, 
\left(\matrix{ 0                       & {\bf m_{_ D}}^{_{\rm T}} \cr
              {\bf m_{_ D}} & {\bf M_{_ R}} \cr }\right),
\label{em6}
\end{equation}
where $3\times 3$ matrices $\bf m_{_D}$ and $\bf M_{_R}$ describe 
the Dirac and
Majorana masses, respectively. For the sake of simplicity we neglect
CP violating phases and assume that $\bf m_{_D}$ and $\bf M_{_R}$ are real.
We investigate a question 
if eq. (\ref{mutaumix}) can be obtained without imposing 
large cancellations and correlations between parameters in 
$\bf M_{_R}$ and  $\bf m_{_D}$.

After diagonalization of the mass matrix (\ref{em6}) the particle spectrum
consists of three light and nearly left-handed neutrinos and three nearly
right-handed neutrinos whose masses are of order of a huge mass scale $M$.
It is assumed that non-zero elements of $\bf M_{_R}$ are of order of $M$
and hence much larger than the matrix elements of $\bf m_{_D}$. The masses
of the light neutrinos $\nu_{_L}$ can be calculated by considering Dirac
masses as small perturbations. In the second order of perturbation theory
the following $3\times 3$ mass matrix is obtained
\begin{equation}
\label{numass}
{\bf N} \, =\,  {\bf m_{_ D}}^{\rm T}\, {\bf M_{_ R}}^{-1}\, {\bf m_{_ D}}.
\end{equation}
The masses of $\nu_{_L}$'s are equal to absolute values of its eigenvalues.
Inversion of the matrix $\bf M_{_R}$ is possible because the condition
\begin{equation}
\label{detMR}
{\rm det}\, {\bf M_{_ R}} \ne 0
\end{equation}
is assumed, which guarantees that all right-handed neutrinos are heavy.

One might argue that eq.~(\ref{mutaumix}) cannot impose any limitation on
models because for an arbitrary non--singular matrix ${\bf N}$, given the 
form of Dirac masses $\bf m_{_D}$, the mass matrix for Majorana masses is
equal to
\begin{equation}
\label{eMR}
{\bf M_{_ R}} \, =\, {\bf m_{_ D}}\, {\bf N}^{-1}\,{\bf m_{_ D}}^{\rm T}.
\end{equation}
Thus, one can choose $\bf M_{_R}$ such that an arbitrary mass and mixing 
pattern is obtained. However, in general eq.~(\ref{eMR}) implies large
correlations between low mass parameters in $\bf m_{_D}$ and large mass
parameters in $\bf M_{_R}$. Such correlations, if not removed by a freedom
in defining $\bf m_{_D}$, are unnatural because $\bf m_{_D}$ and
$\bf M_{_R}$ originate from apparently disconnected mechanism, e.g. from
local gauge symmetry breakings for electroweak $SU_2\times U_1$ and some
grand unification group $G$. Then another rather difficult problem arises
if such correlations are stable against radiative corrections. For these
reasons we consider only those models which imply eq.(\ref{mutaumix})
without fine tuning between the parameters at low and large mass scales.
Let us choose the basis in which the matrix $\bf m_{_D}$ is diagonal
\begin{equation}
{\bf m_{_ D}} = {\rm diag}(m_1,m_2,m_3) = m_3\, {\rm diag}(x^2 y,x,1).
\label{diagonal}
\end{equation}
The condition which we impose means that for acceptable models 
a system of coordinates in generation space exists such that
the Majorana mass matrix $\bf M_{_R}$ does not depend on the parameters
$x$ and $y$.
The sector of Majorana masses depends on some other set of parameters 
$\{ \alpha,\beta, \dots\} $. 
We do not preclude any mass hierarchy at the high
mass scale so some of these parameters may be small. We consider that
those models are unnatural whose essential features, like the patterns of
their mass spectra, depend in a crucial way on relations between the 
parameters describing the low and the high mass sectors. Thus, we do not
discuss models which assume relations like $\alpha/x \approx 1$ or any
other relation implying strong correlations between the parameters in
$\bf m_{_D}$ and $\bf M_{_R}$.

Assuming mass relations typical for $SO_{10}$ grand unified theories,
$\bf m_{_D}$ should follow the mass pattern of up type quarks whereas
$\bf m_{_{\ell}}$, the mass matrix for the charged 
leptons should resemble the corresponding matrix for down type quarks.
Moreover, since the Cabibbo--Kobayashi--Maskawa mixing for the second
and third generation of quarks is quite small, in the present basis
$\bf m_{_{\ell}}$ is `nearly' diagonal. The large mixing (\ref{mutaumix})
between $\nu_\mu$ and $\nu_\tau$ originates from the
form of the neutrino mass matrix ${\bf N}$ because the Dirac mass matrices
for all leptons and quarks are nearly diagonal\footnote{This idea has
been considered in the literature. In particular a classification of some
phenomenologically attractive textures is given in \cite{LLSV}. These 
papers provide also a physical motivation (i.e. a symmetry) for the
corresponding mass matrix ans\"atze.}. 
In the following discussion
we take into account that the mass eigenstates of charged leptons can be 
imperfectly aligned with our coordinates in the generation space. 
Namely the mass matrix for the charged leptons may have small off-diagonal 
elements, whose ratios to the diagonal elements would be comparable to the 
mixing angles in the CKM matrix for quarks.
We write
\begin{equation}
\label{MRinv}
\left({\bf M_{_ R}}^{-1}\right)_{ij} \, =\,  {1\over M}\, {\bf a}_{ij} ,
\end{equation}
where the mass scale $M$ is chosen in such a way that
\begin{equation}
{\rm max}\left(\, \left| a_{ij}\right|\,\right) = 1.
\end{equation}
It follows that 
\begin{equation}
{\bf N}\, =\, {{m_3}^2\over M}\, 
\left(\matrix{ x^4 y^2 a_{11} & x^3 y a_{12} & x^2 y a_{13}  \cr
               x^3 y a_{12}   & x^2 a_{22}   & x a_{23}      \cr
               x^2 y a_{13}   & x a_{23}     & a_{33}        \cr}\right).
\label{Nmatrix}
\end{equation}
A natural order of magnitude for the parameters $x$ and $y$ is
\begin{equation}
x = O\left( m_c/m_t\right)\sim 10^{-2} ,\qquad\qquad
y = O\left( m_u m_t/ m_c^2\right)\sim 10^{-1} .
\end{equation}
Therefore if $a_{33}= O(1)$ no large mixing is possible between the third
and the first two generations. In such a case the mass
$\mu_3$ of the heaviest mass eigenstate $\nu_3$ is much larger 
than $\mu_2$ and $\mu_1$ corresponding to the other two mass 
eigenstates $\nu_2$ and $\nu_1$.
The maximal mixing (\ref{mutaumix}) implies that
\begin{equation}
a_{33} = 0
\label{a33}
\end{equation}
or at least $a_{33}$ has to be strongly suppressed. 
If the condition (\ref{a33}) is fulfilled and
\begin{equation}
a_{23} = O(1) \ne 0
\label{a23}
\end{equation}
the large mixing (\ref{mutaumix}) is obtained. The masses $\mu_3$ and
$\mu_2$ are both of the same order of magnitude
\begin{equation}
\mu_3 \approx \mu_2 \approx 
{{m_3}^2 \over M}\, x \,\left| a_{23}\right| .
\end{equation}
The mass splitting between $\nu_3$ and $\nu_2$ depends on the values of 
matrix elements $a_{22}$ and  $a_{13}$ and is typically of order
$${{m_2}^2 \over M}\, O\left(a_{22}, y a_{13}\right)$$
i.e. suppressed by one power of $x$ with respect to $\mu_3$ and 
$\mu_2$. The mass of the eigenstate $\nu_1$ is typically much smaller 
than $\mu_2$ and the mixing angles $\theta_{e\mu}$ and $\theta_{e\tau}$
are also small.
As an example of this class of models let us consider $a_{12}=a_{13}=0$,
$\left|a_{22}/a_{23}\right|=\alpha$ and  $\left|a_{11}/a_{23}\right|=\beta$.
A large mixing (\ref{mutaumix}) is obtained if $\alpha\ll 1/x$. Then
the following mass spectrum of the light neutrinos is derived:
\begin{eqnarray}
\mu_3 &=& \mu \left( 1 + {\textstyle{1\over 2}}\alpha x + \dots \right)
\nonumber\\
\mu_2 &=& \mu \left( 1 - {\textstyle{1\over 2}}\alpha x + \dots \right)
\nonumber\\
\mu_1 &=& \mu\, \beta x^3 y^2
\label{massesI}
\end{eqnarray}
with\footnote{A model for two nearly degenerate neutrino mass eigenstates 
similar to (\ref{massesI}) for $\alpha\approx 2$
has been recently discussed in \cite{Allanach}; see also \cite{CalMoh}.} 
\begin{equation}
\mu = m_2 m_3 \left|a_{23} \right|/ M .
\end{equation}
It is evident that for $\beta x^3 y^2 \ll 1$, 
\begin{equation}
{\Delta m_{23}^2 \over \Delta m_{12}^2} \approx
{\Delta m_{23}^2 \over \Delta m_{13}^2} = O(x) ,
\label{ratios}
\end{equation}
where
\begin{equation}
\Delta m_{ij}^2 = \left|\, {\mu_i}^2 - {\mu_j}^2 \, \right|.
\end{equation}
The mass scale $\mu$ can be estimated assuming
\begin{equation}
\Delta m_{23}^2 = 2\cdot 10^{-3}\, {\rm eV}^2
\label{nutauSK}
\end{equation}
as obtained by the Super-Kamiokande collaboration \cite{SKam}
and $x \approx 1/100$
\begin{equation}
\mu \approx \sqrt{\Delta m_{23}^2\over 2\alpha x}
\approx {0.3\over\sqrt{\alpha}}\, {\rm eV} .
\label{muest}
\end{equation}
The parameter $\alpha$ in (\ref{muest}) may be small. In such 
a case $\nu_\mu$ and $\nu_\tau$ are pushed even more towards the
maximal mixing. For small $\alpha$ the mass $\mu$ may become 
larger than 1 eV. It is possible in this class of models  
that $\nu_\mu$ and $\nu_\tau$ contribute a non-negligible 
contribution to the dark matter in the Universe. It is also quite
natural to expect neutrino oscillations between $\nu_e$ and $\nu_\mu$
as well as between $\nu_e$ and $\nu_\tau$ characterized by small
mixing angles and, at a fixed energy, by much shorter oscillation
lengths than for $\nu_\mu \to \nu_\tau$, c.f. eq.~(\ref{ratios}). 
In the system of coordinates which we use the mixing angles for quarks 
are ascribed to weak isospin $I_3 = - {1\over 2}$ states.
The Dirac mass matrix for the down type quarks, and hence also 
for the charged leptons is not diagonal. For the mixing between the
first and second generation 
$\theta_{e\mu} \sim \theta_c$ seems quite natural, where $\theta_c$
denotes the Cabibbo angle. 
\par\noindent
It is interesting that a small mixing
angle $\theta_{e\mu}$ as well as $\Delta m_{12}^2$ given by
\begin{equation}
\Delta m_{12}^2 \approx \Delta m_{13}^2 \approx \mu^2
\end{equation}
are well within the range of parameters allowed by $\nu_\mu \to \nu_e$ 
oscillations reported by LSND collaboration \cite{LSND}.
A real drawback of these models is, however, that they evidently
cannot explain deficits of solar neutrinos by a mixing between 
$\nu_e$, $\nu_\mu$ and $\nu_\tau$ only. 

Let us consider now another class of models corresponding to the following
choice of the matrix elements in (\ref{MRinv}):
\begin{equation}
a_{22} = 1, \qquad a_{23} = x, \qquad \qquad a_{33} = x^2.
\label{chytry}
\end{equation}
This choice apparently violates the condition that Majorana mass matrix 
$\bf M_{_R}$ should not depend on parameters in Dirac mass matrix 
$\bf m_{_D}$. However the $x$-dependence of the elements $a_{ij}$
in eq.(\ref{chytry}) can be absorbed into an orthogonal matrix describing
a rotation by a small angle $x$ in the $\nu_\mu - \nu_\tau$ plane 
\begin{equation}
{\bf M_{_ R}}^{-1}(x) = 
{\bf O}^{\rm T}(x)\, {\bf M_{_ R}}^{-1}(x=0)\, {\bf O}(x)
\label{rotation}
\end{equation}
where
\begin{equation}
{\bf O}(x) \approx \left( \matrix{ 1  &  0   &   0  \cr
                             0  &  1   &   x  \cr
                             0  &  -x  &   1  \cr  } \right) .
\end{equation}
Then all $x$-dependence can be ascribed to a new Dirac mass matrix
\begin{equation}
{\bf m_{_ D}^\prime}  =  {\bf O}(x) {\bf m_{_ D}} = m_3\,
              \left( \matrix{x^2y  &  0     &   0  \cr
                             0     &  x     &   x  \cr
                             0     &  -x^2  &   1  \cr  } \right) .
\label{mDprime}
\end{equation}
When rewritten in this way our model is free from dangerous correlations
between the parameters in Dirac and Majorana mass matrices. It does not 
mean that no small parameters are present. In fact $x$ and $y$ have to
be small in a realistic model. Thus we allow also for small parameters,
independent of $x$ and $y$, in the Majorana mass matrix 
${\bf M_{_ R}}^{-1}(x=0)$.
\par\noindent
As a specific example we consider
\begin{equation}
{\bf M_{_ R}}^{-1}(x=0) =  {1\over M}
             \left( \matrix{ 0       &  0   &   \alpha  \cr
                             0       &  1   &   0       \cr
                             \alpha  &  0   &   0       \cr  } \right) .
\label{MR0}
\end{equation}
and obtain the characteristic equation for the eigenvalues of
${\bf N} M/{m_2}^2$.
When small terms of order $x^2$ are neglected this equation
reads:
\begin{equation}
\lambda^3 - 2\lambda^2 - r^2 \lambda + r^2 = 0 ,
\label{chareq}
\end{equation}
where
\begin{equation}
r =\alpha y .
\end{equation}
In the limit of small $r$ the eigenvalues are given by
\begin{eqnarray}
\lambda_1 &=& \; {\textstyle{1\over \sqrt{2}}} r - {\textstyle{1\over 8}} r^2
+ \dots , \nonumber\\
\lambda_2 &=& - {\textstyle{1\over \sqrt{2}}} r - {\textstyle{1\over 8}} r^2
+ \dots ,   \nonumber\\  
\lambda_3 &=& 2 + {\textstyle{1\over 4}} r^2 + \dots ,   
\end{eqnarray}
and the corresponding eigenvectors are proportional to
\begin{equation}
v_1 \sim \,  \pmatrix{ \left.
\matrix{  \sqrt{2} +  r/4 \cr  
- 1  -  r/\sqrt{2}  \cr  1  \cr }\right) + \dots,\quad
v_2 \sim \, \left(
\matrix{ -\sqrt{2} +  r/ 4 \cr 
- 1  -  r/\sqrt{2}  \cr  1  \cr }\right) + \dots,\quad
v_3 \sim \,  \left( 
\matrix{   r/2 \cr   1  \cr  1  \cr }\right. } + \dots
\label{eigenvect}
\end{equation}
The mass spectrum of the nearly left-handed neutrinos $\nu_L$ is
\begin{eqnarray}
\mu_1 &=& \mu\,  r\, \left(
\; {\textstyle{1\over \sqrt{2}}}  - {\textstyle{1\over 8}} r
+ \dots \right)  , \nonumber\\
\mu_2 &=& \mu\, r\,  \left(
 {\textstyle{1\over \sqrt{2}}}  + {\textstyle{1\over 8}} r
+ \dots \right) ,   \nonumber\\  
\mu_3 &=& \mu \,  \left(2 + {\textstyle{1\over 4}} r^2 + \dots \right) 
\label{spectrumII}   
\end{eqnarray}
with
\begin{equation}
\mu = {m_2}^2/M  .
\end{equation}
Thus two almost degenerate mass eigenstates $\nu_1$ and $\nu_2$
corresponding to the eigenvalues $\mu_1$ and $\mu_2$ are lighter 
than the third eigenstate $\nu_3$ whose mass is greater by a factor
${2\sqrt{2}/ r}$. 
 
In the leading order 
of the small parameter $r$ the eigenstates of the weak charged current are 
expressed in terms of the mass eigenstates in the following way
\begin{eqnarray}
\ket{\nu_e} &=& {\textstyle{1\over \sqrt{2}}} 
\left[\  \ket{\nu_1} - \ket{\nu_2}\, \right] + \dots, \nonumber\\
\ket{\nu_\mu} &=& {\textstyle{1\over \sqrt{2}}} \left[ \, \ket{\nu_3}
      - {\textstyle{1\over \sqrt{2}}}
\left( \, \ket{\nu_1} + \ket{\nu_2} \,\right) \, \right]
+ \dots, \nonumber\\
\ket{\nu_\tau} &=& {\textstyle{1\over \sqrt{2}}} \left[ \, \ket{\nu_3}
      + {\textstyle{1\over \sqrt{2}}}
\left(\,  \ket{\nu_1} + \ket{\nu_2}\, \right) \, \right]
+ \dots ,
\label{bi-maximal}
\end{eqnarray}
where $\dots$ denote terms of order $r$ and smaller.
The oscillations of $\nu_\mu$ observed by the Super-Kamiokande collaboration
can be understood as an oscillation of the states between $\ket{\nu_\mu}$ 
and $\ket{\nu_\tau}$. 
Since the eigenmasses $\mu_1$ and $\mu_2$ are degenerate to a high 
degree there will be no relative phase change between $\ket{\nu_1}$ 
and $\ket{\nu_2}$ while travelling the distance of the Earth size
because $\Delta m_{12}^2/E \ll 1/R_{Earth}$. Meanwhile, the splitting
between $\mu_3$ and $\mu_1\approx\mu_2$ allows for a significant phase
difference between $\ket{\nu_3}$  and 
$\left[\,\ket{\nu_1} + \ket{\nu_2}\,\right]/\sqrt{2}$ 
after time $t\sim R_{Earth}$ of neutrino propagation through the Earth.
This implies a significant oscillation between $\nu_\mu$ and $\nu_\tau$
at the scale of $R_{Earth}$ if $\Delta m_{23}^2/E \sim 1/R_{Earth}$.
At such a scale effectively the maximal mixing is observed.
As for solar
neutrinos the oscillations can be understood as a result of time
evolution of the state vector in a two-dimensional vector space spanned
by $\ket{\nu_1}$ and $\ket{\nu_2}$. As a result of being 
nearly orthogonal to $\ket{\nu_e}$ the mass eigenstate $\ket{\nu_3}$
remains nearly orthogonal to the state vector which
during its time evolution oscillates between $\ket{\nu_e}$ and
$\left[\,\ket{\nu_\mu} - \ket{\nu_\tau}\,\right]/\sqrt{2}$.
It is remarkable that a maximal mixing for oscillations of electron
neutrinos is predicted by the model, exactly as needed for explaining the
solar neutrino problem by vacuum oscillations \cite{just-so}. 
Numerical values of $\mu$ and $r$ can be derived from the Super-Kamiokande
result (\ref{nutauSK}) and the value of $\Delta m_{12}^2$ obtained
from quantitative analyses of solar neutrino oscillations in vacuum
\cite{FY}. It follows from the mass spectrum (\ref{spectrumII}) that
\begin{equation}
\Delta m_{13}^2 \approx \Delta m_{23}^2 \approx 4 \mu^2 
\approx 2\cdot 10^{-3} \ {\rm eV}^2 ,
\end{equation}
\begin{equation}
\Delta m_{12}^2 \approx {\mu^2 r^3\over 2\sqrt{2}} \sim 10^{-10}\ 
{\rm eV}^2.
\label{Deltam12}
\end{equation}
Taking central values one obtains $r = 0.008$ and $\mu = 0.02$~eV.
Assuming that $m_2$ is equal to the mass of the charm quark the scale of
the Majorana mass sector M is between $10^{10}$ and $10^{11}$~GeV 
depending on the scale at which $m_c$ is evaluated.
\vskip0.3cm
\par\noindent
{\bf Note added:}
After this paper had been submitted for publication we learned about 
recent works by V. Barger, S. Pakvasa, T.J. Weiler and K. Whisnant
\cite{BPWW} and by A.J. Baltz, A.S. Goldhaber and M. Goldhaber \cite{BGG}.
In \cite{BPWW,BGG} the so-called bi-maximal mixing has been proposed
assuming on phenomenological grounds the relation (\ref{bi-maximal}) 
between the mass and gauge eigenstates. Phenomenological consequences
of such a scenario are discussed therein. The present work shows that
the bi-maximal mixing can be derived in see-saw models without arranging
correlations between parameters in the Dirac and Majorana mass sectors.
The second model (\ref{mDprime})--(\ref{MR0}) leads to a characteristic mass
and mixing pattern (\ref{spectrumII})--(\ref{bi-maximal}) for the
light neutrinos. This pattern remains practically unchanged when the
three elements in the left--upper corner of the matrix (\ref{MR0})
are non-zero and of order one. 
Our model is a particular realization
of the bi-maximal mixing scenario because not only
the values of $\Delta m_{ij}^2$ but also all the masses 
are specified. The value of the parameter 
$r$ is not much changed when a recent fit to the solar neutrino data
\cite{Bahcall} is used implying $\Delta m_{12}^2 \approx 
7\cdot 10^{-11}\ eV^2$
instead of eq.(\ref{Deltam12}). However, it changes significantly in
the range of $\Delta m_{12}^2$ considered in \cite{BGG}. 
There is no neutrinoless double beta decay or at least 
the rate of this process is strongly suppressed for $a_{11}\ne 0$
because the element ${\bf N}_{11}$ of the neutrino mass matrix (\ref{Nmatrix})
is much smaller than the masses $\mu_1$ and $\mu_2$: 
${\bf N}_{11}= \mu x^2 y^2 a_{11}$.
Our model shows that the mass parameter describing neutrinoless double
beta decay can be much smaller than the Majorana masses of neutrino mass 
eigenstates.

\section*{\bf Acknowledgements}
We would like to thank Hans K\"uhn for his warm hospitality creating 
a very nice and stimulating atmosphere in the Institut f\"ur Theoretische 
Teilchenphysik, Universit\"at Karlsruhe.\\
Helpful discussions with Wolfgang Hollik, Thomas Mannel and Michael
G. Schmidt are gratefully acknowledged.\\
This work is partly supported by KBN grant 2P03B08414, by BMBF grant
POL-239-96 and by the Alexander von Humboldt Foundation.


\begin{thebibliography}{99}
\bibitem{SKam}
Super--Kamiokande Collaboration., Y. Fukuda et al., Phys. Lett. {\bf B433}
(1998) 9; hep-ex/9805006 and hep-ex/9807003; 
talk by Y. Kajita  at Neutrino--98, Takayama, Japan, June 1998. 
\bibitem{YGMRS}
T. Yanagida , in: O. Sawada and A. Sugamote (eds.), {\it Proc. of the Workshop
on the Unified Theory and Baryon Number in the Universe} (KEK report 79-18,
1979) p.95;\\ 
M. Gell-Mann, P. Ramond and R. Slansky, in: P. van Nieuwenhuizen and
D.Z. Freedman (eds.), {\it Supergravity} (North Holland, Amsterdam, 1979)
p.315~;\\
S. Weinberg, Phys. Rev. Lett. {\bf 43} (1979) 1566;\\
R.N. Mohapatra and G. Senjanovic, Phys. Rev. Lett. {\bf 44} (1980) 912.
\bibitem{FY}
M. Fukugita and T. Yanagida, in: M. Fukugita and A. Suzuki (eds.),
{\it Physics and Astrophysics of Neutrinos} (Springer--Verlag, Tokyo, 1994)
p.1~. 
\bibitem{LLSV}
G.K. Leontaris, S. Lola, C. Scheich and J.D. Vergados, Phys. Rev. {\bf D53}
(1996) 6381;\\
P. Binetruy, S. Lavignac, S. Petcov and P. Ramond, Nucl. Phys. {\bf B496}
(1997) 3.
\bibitem{Allanach}
B.C. Allanach, hep-ph/9806294.
\bibitem{CalMoh}
D.O. Caldwell and R.N. Mohapatra, Phys. Rev. {\bf D48} (1993) 3259.
\bibitem{LSND}
C. Athanassopoulos et al. (LSND Collab.),UCRHEP-E197 and nucl-ex/9709006.
\bibitem{just-so}
J.G. Learned, S. Pakvasa and T.J. Weiler, Phys. Lett. {\bf B207} (1988) 79;\\
V. Barger and K. Whisnant, Phys. Lett. {\bf B209} (1988) 365;\\
K. Hidaka, M. Honda and S. Midorikawa, Phys. Rev. Lett. {\bf 61} (1988) 1537.
\bibitem{BPWW}
V. Barger, S. Pakvasa, T.J. Weiler and K. Whisnant, hep-ph/9806387 and
Phys. Lett. {\bf B} in press.
\bibitem{BGG}
A.J. Baltz, A.S. Goldhaber and M. Goldhaber, hep-ph/9806540.
\bibitem{Bahcall}
J.N. Bahcall, P.I. Krastev and A.Yu. Smirnov, hep-ph/9807216.
\end{thebibliography}
\end{document}